\documentclass[twocolumn,prd,amsmath,showpacs,amssymb]{revtex4}
\usepackage[latin1]{inputenc}
\usepackage{bbm}
\usepackage{bm}
\usepackage{graphicx}
\usepackage{epsfig}
\usepackage{subfigure}
\usepackage{latexsym}
\usepackage{amsmath}
\usepackage{amsfonts}
\usepackage{amssymb}
\usepackage{color}

\newcommand{\Dslash}{\ensuremath \raisebox{0.025cm}{\slash}\hspace{-0.28cm} D}

\def\tr{\mathrm{Tr}}

\newcommand{\be}{\begin{equation}}
\newcommand{\ee}{\end{equation}}

\newcommand{\bea}{\begin{eqnarray}}
\newcommand{\eea}{\end{eqnarray}}

\begin{document}
\title{Naive Dimensional Analysis and Irrelevant Operators}
\author{Matti Antola$^{a,c}$}
\author{Kimmo Tuominen$^{b,c}$}
\affiliation{$^a$
Department of Physics, 
P.O.Box 64, FI-000140, University of Helsinki, Finland}
\affiliation{$^b$
Department of Physics, 
P.O. Box 35, FI-40014 University of Jyv\"askyl\"a}
\affiliation{$^c$
Helsinki Institute of Physics, 
P.O.Box 64, FI-000140, University of Helsinki, Finland}

\begin{abstract}
We derive a set of easy rules to follow when estimating the coefficients of operators in an effective Lagrangian. In particular, we emphasize how to estimate the size of the coefficients originating from irrelevant interactions in the underlying Lagrangian.
\end{abstract}

\maketitle

Effective field theory constitutes an essential method in modern theoretical physics, see e.g. \cite{Kaplan:2006sv}. In  elementary particle physics some of the well known examples are heavy quark effective theory \cite{Manohar:2000dt} and, likely, the entire Standard Model itself. In typical applications of effective theory one constructs the model Lagrangian based only on the knowledge of the global symmetries and relevant degrees of freedom. Then, in the absence of matching onto a more microscopic theory (or data), it is important to be able to estimate the size of the dimensionless coefficients of the terms arising in the effective Lagrangian.

For illustration, consider QCD with a four-fermion interaction
\be
\mathcal{L}= \overline{Q}\left(i\,\,\Dslash -m \right)Q+G(\overline{Q}_LQ_R)(\overline{Q}_RQ_L)
\label{NJLlagrange}
\ee
Here $Q$ is an $n_f$ component vector, $m$ is the diagonal matrix of masses and $Q_{L,R}=P_{L,R}Q$ in terms of the usual projectors $P_{L,R}=(1\mp \gamma_5)/2$. We assume that the theory becomes strongly interacting, confines and breaks the chiral symmetry spontaneously according to the known pattern at the scale 
$\Lambda \ll 1/\sqrt{G}$. The dynamical degrees of freedom at low energies are the pions, represented via the matrix $U=\exp(i\pi/f)$, where $f$ is the Goldstone boson decay constant. Their low energy effective theory contains the kinetic theory given by the lowest order chiral perturbation theory and the mass and interaction terms constructed with the spurion method:
\bea
\mathcal{L}_{\rm{eff}}&=& c_0\Lambda^2\tr\left[\partial_\mu U\partial_\mu U^\dagger\right] +c_1 \Lambda^3\tr\left[m(U+U^\dagger)\right] \nonumber\\
&&+c_2\Lambda^6G\tr\left[U\right]\tr\left[U^\dagger\right].\;\;
\label{chipert}
\eea
The mass scale of light non-Goldstone states, $\Lambda$, is the cutoff in the effective theory and also the matching scale between the effective Lagrangian and the high energy Lagrangian. We write this scale explicitly in the effective Lagrangian to make the a priori unknown coefficients $c_i$ dimensionless. The task now is to estimate these coefficients, and capture the relative importance of the terms in the effective Lagrangian.

From now on we will work in units where $\Lambda=1$. We also ignore any $\mathcal{O}(1)$ factors, i.e. we only keep track of powers of $\Lambda/f\equiv g$ and coupling constants.

The first constant, $c_0$, can be fixed by requiring the pion fields to have a canonical kinetic term. This leads to $c_0=1/g^2$. To estimate the constants $c_1$ and $c_2$, we need other methods.

In QCD a robust prediction is made by naive dimensional analysis (NDA) \cite{Manohar:1983md}, which was generalized by Georgi \cite{Georgi:1992dw} to apply to more general theories than QCD. According to Georgi, the coefficients of an operator in an effective Lagrangian for a strongly interacting theory depend on $\Lambda$ and $g\equiv\Lambda/f$ in the following simple way:
\begin{enumerate}
\item Divide each term by $g^2$
\item Each strongly interacting field is accompanied by a factor of $g$
\item Fix the overall dimension by multiplying with $\Lambda$
\end{enumerate}
The item two in the above list is taken into account automatically by writing $U=\exp(ig\pi)$. With this in mind, application of the rules yields $c_1=c_2=1/g^2$, and hence the generalized NDA results in
\be
m_\pi^2\sim m+G.
\label{resGNDA}
\ee
However, from the underlying Lagrangian (\ref{NJLlagrange}) we expect
\be
m_\pi^2\sim mg^2\left<\overline{Q}Q\right>+Gg^2\left<\overline{Q}Q\right>^2,
\label{QCDlow}
\ee
and comparing with (\ref{resGNDA}) we see that the correct dependency on $g$ in both terms cannot be obtained with generalized NDA: either $\left<\overline{Q}Q\right>\sim 1/g^2$ or $\left<\overline{Q}Q\right>\sim 1/g$, but in both cases the resulting total dependence on $g$ is wrong. Since numerically in QCD  $\left<\overline{Q}Q\right>\sim \Lambda^3/g^2$ we assume the first is more correct, and that the rules of generalized NDA should be improved to properly take into account the suppression of irrelevant operators. In this paper we derive such rules.

The basic assumption of NDA is that in a strongly interacting system, diagrams of each loop order give the same size contributions. From momentum conservation the number of loops in a diagram is $L=P-V+1$, where $P$ is the number of propagators and $V$ is the number of vertices. In units $\Lambda=1$, each loop is associated with a factor $1/16\pi^2$. Then, if the propagator is of the order $\alpha$ and all vertices are assumed to be of similar size, $\beta$, the following should hold:
\be
\beta=\frac{1}{(16\pi^2)^{L}}\alpha^P\beta^V=\frac{1}{16\pi^2}
\left(\frac{\alpha}{16\pi^2}\right)^P\left(16\pi^2\beta \right)^V\;.
\ee
The solution valid for any $P$ and $V$ is $\beta=1/\alpha=1/16\pi^2$. Therefore NDA requires that in some non canonical field normalization, each operator coefficient is $\mathcal{O}(1/16\pi^2)$.

We generalize and explicate this statement in two ways. First, we write the factor $1/16\pi^2$ as $1/g^2$, as was already done by Georgi \cite{Georgi:1992dw}. By allowing $g\lesssim 4\pi$ one can include effects of small parameters present in e.g. large-N field theories. Second, we apply these principles at the matching scale $\Lambda$ to both the fundamental and effective Lagrangians, also including any explicit symmetry breaking or irrelevant operators, as was done in \cite{Cohen:1997rt,Luty:1997fk}.

More specifically, let $\phi$ and $M$ collectively denote all the strongly interacting fields present in the high and low energy Lagrangian, and let the hatted fields denote corresponding non canonically normalized fields. Then the principles above are equivalent to the matching
\be
\mathcal{L}(\phi)=\frac{1}{g^2}\hat{\mathcal{L}}(\hat{\phi})\;\;\rightarrow\;\;
\mathcal{L}_{\textrm{eff}}(M)=\frac{1}{g^2}\hat{\mathcal{L}}_{\textrm{eff}}(\hat{M})\;.
\label{generalmatching}
\ee
where the hatted Lagrangians include only $\mathcal{O}(1)$ coefficients.

To clarify the idea, we now reconsider the previous example. We require that the fundamental Lagrangian (\ref{NJLlagrange}) takes the form
\be
\mathcal{L}= \frac{1}{g^2}\left[\overline{\hat{Q}}\left(i\,\,\Dslash -1\right)\hat{Q}+
(\overline{\hat{Q}}_L\hat{Q}_R)(\overline{\hat{Q}}_R\hat{Q}_L)\right]\;.
\ee
This matches with (\ref{NJLlagrange}) if
\be
\hat{Q}=gQ\;\;,\;\;m=I\;\;,\;\;G=g^2\;.\label{match0}
\ee
Now we apply NDA to the effective Lagrangian as well,
\bea
\mathcal{L}_{\textrm{eff}}&=& \frac{1}{g^2}\left(\tr\left[\partial_\mu \hat{U}\partial_\mu \hat{U}^\dagger\right]+\tr\left[(\hat{U}+\hat{U}^\dagger)\right]  \right. \nonumber \\
&&+\left.\tr\left[\hat{U}\right]\tr\left[\hat{U}^\dagger\right]\right),
\label{leff}
\eea
where $\hat{U}=\exp{i\hat{\pi}}$. To restore the usual kinetic term for $\pi$ we expand the exponential and find that $\hat{\pi}=g\pi$ and $U=\exp{ig\pi}$.

We now reinstate the spurion parameters $m$ and $G$. In the limit these parameters are zero, the second and third terms in (\ref{leff}) vanish by symmetry arguments, so the coefficients must be proportional to $m$ and $G$. The proportionality is fixed by requiring that in the limit $m=I$ and $G=g^2$, one should obtain (\ref{leff}). Therefore
\bea
\mathcal{L}_{\textrm{eff}}&=&\frac{1}{g^2}\left(\tr\left[\partial_\mu \hat{U}\partial_\mu\hat{U}^\dagger\right]+\tr\left[m(\hat{U}+\hat{U}^\dagger)\right]  \right. \nonumber \\
&&
+\left.\frac{G}{g^2}\tr\left[\hat{U}\right]\tr\left[\hat{U}^\dagger\right]\right)\label{result}\\
&=& \frac{1}{g^2}\tr\left[\partial_\mu U\partial_\mu U^\dagger\right]+\frac{1}{g^2}\tr\left[m(U+U^\dagger)\right]\nonumber \\
&&  +\frac{G}{g^4}\tr\left[U\right]\tr\left[U^\dagger\right].
\eea
Hence we obtain $c_1=1/g^2$, but $c_2=1/g^4$, and
\be
m_\pi^2\sim m+\frac{G}{g^2}.
\label{mpioni}
\ee
Thus we find behavior compatible with (\ref{QCDlow}) if $\left<\overline{Q}Q\right>\sim 1/g^2$.

As an outcome of this treatment, the irrelevant interactions in the underlying Lagrangian have become more suppressed in the effective Lagrangian than in generalized NDA. Also, we find the maximum values of the spurion parameters for which the effective Lagrangian makes sense: clearly, the elements of the matrix $m$ should be less than one, and $G<g^2$; otherwise terms of higher order in the spurions will be larger than these lowest order terms.

Our result is based on a simple scaling argument, and, as already noted, similar considerations have appeared in literature \cite{Cohen:1997rt, Luty:1997fk}. However, one would expect a general description akin to the rules of the generalized NDA, and we are not aware that they have been presented elsewhere. Therefore we now generalize the above discussion and derive simple rules for applications and also compare to the rules of NDA presented in \cite{Georgi:1992dw}.

In the notation of (\ref{generalmatching}), it is difficult to account for operator coefficients such as $G$, which might be suppressed by additional powers of $g$ in the effective Lagrangian. It is also difficult to account for any dependence on $g$ in the high energy Lagrangian. We thus rewrite the condition in two steps. In step one we require (\ref{generalmatching}) for the kinetic terms, and in step two we generalize (\ref{generalmatching}) for other terms.

Let the canonical kinetic terms in the high and low energy Lagrangians be $\mathcal{O}_{\textrm{KE}}(\phi,g)$ and $\mathcal{O}_{\textrm{eff,KE}}(M,g)$, where we have explicated the dependence on $g$ and fields. We assume that the kinetic energy operators do not depend on parameters other than $g$. Therefore $\mathcal{O}_{\textrm{KE}}(\phi,1)$ and $\mathcal{O}_{\textrm{eff,KE}}(M,1)$ have $\mathcal{O}(1)$ coefficients, and step one is to require
\bea
\mathcal{O}_{\textrm{KE}}(\phi,g)
&=&\frac{1}{g^2}\mathcal{O}_{\textrm{KE}}(\hat\phi,1)
\nonumber\\
\mathcal{O}_{\textrm{eff,KE}}(M,g)
&=&\frac{1}{g^2}\mathcal{O}_{\textrm{eff,KE}}(\hat M,1). \label{matchingke}
\eea
From these equations we can solve $\phi(\hat\phi,g)$ and $\hat M(M,g)$. We now assume these relations are known. 

Step two of the new matching is to require
\be
\mathcal{L}(\phi,g)=\frac{1}{g^2}\hat{\mathcal{L}}(\hat{\phi},g)\;\;\rightarrow\;\;
\mathcal{L}_{\textrm{eff}}(M,g)=\frac{1}{g^2}\hat{\mathcal{L}}_{\textrm{eff}}(\hat{M},g)\;.
\label{matching3}
\ee
where now instead of requiring that $\hat{\mathcal{L}}(\hat{\phi})$ and $\hat{\mathcal{L}}_{\rm{eff}}(\hat{M})$ have $\mathcal{O}(1)$ coefficients, we require that $\hat{\mathcal{L}}_{\textrm{eff}}(\hat{M},g)$ is built with operators that have the same dependence on $g$ and other couplings as operators in $\hat{\mathcal{L}}(\hat{\phi},g)$. 

To clarify this condition, consider a general operator $\mathcal{O}(\phi,g)$ in $\mathcal{L}$ that can contain a small parameter. The high energy matching condition in (\ref{matching3}) becomes
\be
\mathcal{O}(\phi(\hat{\phi},g),g)=\frac{1}{g^2}\widehat{\mathcal{O}}(\hat{\phi},g)\;.\label{hmatching}
\ee
This equation defines $\widehat{\mathcal{O}}(\hat{\phi},g)$, and especially its dependence on $g$.

Now we will transition to the effective Lagrangian. It is written in terms of effective operators 
\be
\widehat{\mathcal{O}}_\mathrm{eff}(\hat{M}(\hat\phi),g)\sim\widehat{\mathcal{O}}(\hat{\phi},g)\;.
\label{Ocorr}
\ee
The correspondence means that the symmetry breaking structure, dependence on couplings, and dependence on $g$, is the same for both operators. From the low energy condition in (\ref{matching3}) we directly find that any term in the effective Lagrangian is of the form
\be
\mathcal{L}_\textrm{eff}(M,g)\supset\frac{1}{g^2}\left(\widehat{\mathcal{O}}_\mathrm{eff}(\hat{M}(M,g),g)\right)^n\;.\label{effn}
\ee

The generalization to many operators $\mathcal{O}_i$ is straightforward. However, now we require that all fields are in a linear representation and all operators are polynomial in the fields. Therefore we write $\mathcal{O}_i(\phi,g)\equiv c_i(g)h_i(\phi)$ where all the dependence on $g$ and other couplings is in the coefficient $c_i(g)$. We write a general Lagrangian,
\be
\mathcal{L}(\phi,g)=\sum_k c_k h_k(\phi)\;,
\ee
where the operators $c_k h_k(\phi)$ enumerate all terms in the Lagrangian and we have omitted the dependence on $g$ in the coefficient.

We first solve the field rescaling $\hat\phi(\phi,g)$ and $\hat M(M,g)$ from (\ref{matchingke}). For any type of field (scalar, gauge, or fermion), the kinetic term is comprised of two fields so we find $\hat{\phi}=g\phi$ and $\hat M=gM$.

Next we apply the high energy condition (\ref{matching3}). If $s_k$ gives the number of strongly interacting fields in the operator $h_k(\phi)$, we find
\be
c_k h_k(\phi)=\frac{1}{g^2} \hat{c}_k h_k(\hat\phi)=g^{s_k-2}\hat{c}_k h_k(\phi)\;,
\ee
where we used the fact that $h_k$ is polynomial in $\phi$ for the second equivalence. This equation is solved for $\hat c_k=g^{-\chi_k}c_k$ where $\chi_k=s_k-2$.

Continuing, according to (\ref{Ocorr}) and (\ref{effn}) the effective Lagrangian is built from operators corresponding to $\hat c_k h_k(\hat \phi)$, and is therefore of the form
\bea
\mathcal{L}_{\textrm{eff}}(M,g)
&=& \frac{1}{g^2}\hat{\mathcal{L}}_{\textrm{eff}}(\hat{M},g)
= \frac{1}{g^2}\left(\sum \hat{c}_{k} h_{k,m}(\hat M)\right.\nonumber \\
&+&
 \left.
 \sum\hat{c}_{k}\hat{c}_{k'}h_{k,m}(\hat M)h_{k',m'}(\hat M)
+...\right)\label{eff}
\eea
Here the new index $m$ in $h_{k,m}(M)$ labels all the possible operators with the same symmetry as $h_k(\phi)$, and in this equation the sum is over all indices $m$, $k$, $m'$, and $k'$.

The last step is to write the operators $h_{k,m}(\hat M)$ of the low energy effective Lagrangian in terms of the hatless fields. Since these operators are also polynomial in the fields we find
\be
h_{k,m}(\hat M)=g^{n_{k,m}}{h}_{k,m}(M),
\ee
where $n_{k,m}$ gives the number of strongly interacting effective fields in the operator.

Combining these results allows us to explicate the powers of $g$ in each term of (\ref{eff}). The result is
\bea
\mathcal{L}_{\rm{eff}} &=& \sum _{k,m} g^{n_{k,m}-\chi_k-2} c_kh_{k,m}\nonumber \\
&+&\sum_{k,m,k',m'}g^{n_{k,m}+n_{k',m'}-\chi_k-\chi_{k'}-2}c_kh_{k,m}c_{k'}h_{k',m'}\nonumber\\
&+&...
\eea
where $h_{k,m}\equiv h_{k,m}(M)$.
 
Now we have access to all elements required to estimate the coefficient of each term arising in the effective Lagrangian. The result can be encapsulated into a form of simple rules which are as follows:
\begin{enumerate}
\item Divide each term by $g^2$
\item Each coupling is accompanied by a factor of $g^{2-n}$, where $n$ is the number of strongly interacting fields in the corresponding high energy operator
\item Each strongly interacting field is accompanied by a factor of $g$
\item Fix the overall dimension by multiplying with $\Lambda$
\end{enumerate}

The only difference to Georgi's rules is the addition of item two. In QCD applications of NDA the high energy Lagrangian consists of the kinetic term and the mass term. Incidentally, for these operators $n-2=0$ in item two above, and both analyses give the same result at each order in the spurions. For terms with more than two strongly interacting fields, like the NJL term which we exhibited in the introduction, the two analyses will give different results.

It should be noted, that if one uses the nonlinear representation for the Goldstone boson matrix $U$, the item three in the rules is taken into account by writing $U=\exp (ig\pi)$.

Also we found that the effective Lagrangian is effectively an expansion in the parameters $\hat c_k$. These should be less than one for the expansion to make sense. For example, in chiral perturbation theory with quark masses, $\hat c_{m_Q}=m_Q/\Lambda$ where $m_Q$ is the mass of the quark and $\Lambda\approx 1$ GeV.

In this paper we have reconsidered NDA. Our results are compatible 
with the ones obtained earlier in e.g. \cite{Cohen:1997rt, Luty:1997fk}, our new output is the derivation of a general set of simple rules which are also able to account for irrelevant interactions. Our results are fully general and can be applied to supersymmetric as well as nonsupersymmetric field theories. We expect our results to be useful in the construction of effective field theories and their application in various contexts.

\end{document}